\documentclass[twocolumn]{bmcart}

\usepackage{amsthm, amsmath}
\usepackage{amssymb} 
\usepackage[pdftex]{graphicx}
\usepackage[caption=false]{subfig}

\RequirePackage[authoryear]{natbib}
\RequirePackage{hyperref}
\usepackage[utf8]{inputenc} 

\startlocaldefs

\newcommand{\vect}[1]{\boldsymbol{#1}}
\newcommand{\MSun}{\ensuremath{M_{\odot}}}
\newcommand{\RSun}{\ensuremath{R_{\odot}}}

\newcommand{\AMUSE}{\texttt{AMUSE}}
\newcommand{\MESA}{\texttt{MESA}}

\endlocaldefs

\begin{document}

\begin{frontmatter}

\begin{fmbox}
\dochead{Research}

\title{Simulations of stripped core-collapse supernovae in close binaries}

\author[
   addressref={aff1},                   
   corref={aff1},                       
   email={rimoldi@strw.leidenuniv.nl}   
]{\inits{AR}\fnm{Alex} \snm{Rimoldi}}
\author[
   addressref={aff1},
   email={spz@strw.leidenuniv.nl}
]{\inits{SPZ}\fnm{Simon} \snm{Portegies Zwart}}
\author[
   addressref={aff1},
   email={emr@strw.leidenuniv.nl}
]{\inits{EMR}\fnm{Elena Maria} \snm{Rossi}}
\address[id=aff1]{
  \orgname{Leiden Observatory, Leiden University},
  \street{Niels Bohrweg 2},
  \city{2333 CA Leiden},
  \cny{Netherlands}
}

\begin{abstractbox}
\begin{abstract}
We perform smoothed-particle hydrodynamical simulations of the explosion of a helium star in a close binary system, and study the effects of the explosion on the companion star as well as the effect of the presence of the companion on the supernova remnant.
By simulating the mechanism of the supernova from just after core bounce until the remnant shell passes the stellar companion, we are able to separate the various effects leading to the final system parameters.
In the final system, we measure the mass stripping and ablation from, and the velocity kick imparted to, the companion star, as well as the structure of the supernova shell.
The presence of the companion star produces a conical cavity in the expanding supernova remnant, and loss of material from the companion causes the supernova remnant to be more metal-rich on one side and more hydrogen-rich (from the companion material) around the cavity.
Following the removal of mass from the companion, we study its subsequent evolution and compare it with a single star not subjected to a supernova impact.
\end{abstract}
\end{abstractbox}

\end{fmbox}

\end{frontmatter}

\section{Introduction} \label{sec:introduction}
There is substantial evidence that most massive stars evolve in binary systems \citep{Duqu91, Rast10, Sana12}.
Therefore, the presence of companion star is an important consideration in the theory and observation of supernovae (SNe) and supernova remnants (SNRs).
In particular, while Type Ia (white-dwarf; WD) supernovae may have a companion which has deposited sufficient mass onto the WD to trigger a `single-degenerate' explosion, many Type Ib/c (stripped core-collapse) supernovae may have close companions that have been at least partly responsible for the loss of mass from the progenitor \citep{Bersten14, Fremling14, Eldridge15, Kim15, Kuncarayakti15}.

Observational searches for supernova companions have typically focused on Type Ia explosions.
Possible companions have been a subject of scrutiny in order to determine the frequency of the two main suspected (single-degenerate or double-degenerate) explosion channels \citep{Nelemans14}.
Hydrogen enrichment from a companion has been searched for in Type Ia SNRs, but so far there has been no evidence of hydrogen lines \citep{Mattila05, Leonard07, Lundqvist15}.
As noted in \cite{Garcia-Senz12}, detection of $H_\alpha$ lines may be difficult due to confusion with Fe and Co lines due to the mostly slow ($< 10^{3}~\mathrm{km\,s}^{-1}$) hydrogen mixing with iron-peak elements.

The presence of a supernova companion is difficult to directly detect if they are low-mass stars at very large distances, and so far definitive evidence of close companions to any SN progenitor, let alone those of Type Ib/c, has been lacking.
Tycho G is probably the best example of a directly imaged, suspected companion, associated with the galactic Type Ia SN, Tycho \citep[SN
1572;][]{Ruiz-Lapuente04}, though recent observations put its status as a supernova companion into dispute \citep{Kerzendorf13, Xue15}.
On the other hand, some direct searches for single-degenerate companions have ruled out giant/subgiant (evolved) stars \citep[SN 2011fe and SNR 1006;][]{Li11, GonzalezHernandez12} and even main-sequence companions \citep[SNR 0509-67.5;][]{Schaefer12}.

The presence of a companion due to increased emission, and therefore modification of the standard light curve, from the ejecta interacting with the companion has also been ruled out in observations of Type Ia supernovae \citep{Olling15}.
However, a recently observed supernova \citep[iPTF14atg;][]{Cao15} does show evidence of interaction with a companion through the detection of an ultraviolet burst in the first several days.
Though much of the focus of previous work has been on Type Ia explosions, the phenomena of companion interactions with single-degenerate Type Ia ejecta has parallels with core-collapse supernovae in binaries, and therefore this scenario still provides a useful context.

\citet[][hereafter, TT98]{TT98} analytically investigated the consequences of a SN in a close, circularised binary, with the goal of finding the runaway velocities of the components of a binary disrupted by a Type Ib/c SN.
This work was calibrated by estimations from early simulations of the effect of a SN shell impact on a star \citep{Fryxell81} in order to determine the amount of mass lost and the change in velocity of the companion.
Motivated by this problem, we perform simulations of SNe in binary systems with properties comparable to those used in TT98.

Simulations of supernovae have been performed at many scales, investigating regions of hundreds of kilometres around the nascent neutron star in the early supernova evolution \citep[for a review of recent progress, see][]{Janka12} to effects of the impact on a companion, or the influence of the companion on the overall structure of the ejecta.
The impact of Type Ia ejecta on companions has, in particular, been well investigated \citep{Marietta00, Pakmor08, Liu12, Pan12}.
\cite{Hirai14} investigated the fraction of mass stripped from a giant companion star due to a effect of a core-collapse (Type II) SN using a two-dimensional grid-based Eulerian code.
Recently, \cite{Liu15} also presented results on the consequences of a Type Ib/c supernova interacting with a binary companion using smoothed-particle hydrodynamics (SPH).
These studies, however, have often focused on the companion star without following the explosion self-consistently from the moment of the supernova.
As a consequence, the ejecta shell is initialised artificially, via analytic prescriptions, near the surface of the companion, without considering its earlier evolution.
In addition, the response of the binary companion, and subsequent supernova remnant evolution, is analysed in these cases from a static configuration rather than placing the binary in an orbit.

For close binary orbits it is typically assumed that the binary has circularised by this point in its evolution, so that the eccentricity
of the orbit can be set to zero (TT98).
We follow the same assumption in this work.
Moreover, despite these close separations and, therefore, short orbital periods, in theoretical work the binary period is taken to be much shorter than the timescale over which the ejected shell impacts the companion.
This can be made more explicit \citep[as in, for example,][]{Colgate70} by noting that the ejecta velocity must be larger than the escape velocity of the primary star,
\begin{equation}
v_\mathrm{ej} > v_\mathrm{esc} = \sqrt{\frac{2 G M_1}{R_1}} ,
\end{equation}
where $M_1$ and $R_1$ are the mass and radius of the primary.
Since the distance $a$ of the companion from the primary is larger than $R_1$, and since the orbital velocity at that distance is
\begin{equation}
 v_\mathrm{orb} = \sqrt{\frac{G M_1}{a}} ,
\end{equation}
then it must be that $v_\mathrm{ej} > v_\mathrm{orb}$.
In practice, the typical ejecta velocities ($10^{3} \sim 10^{4}~\mathrm{km\,s}^{-1}$) are much larger than the orbital velocities
($\sim 10^2~\mathrm{km\,s}^{-1}$), hence the latter is typically ignored in analytic velocity calculations.
However, matter in the ejecta in fact have a radially dependent velocity (approximately, in the homologous regime, $v_\mathrm{ej}(R, t) \propto R/t$).
Therefore, although the density also varies through the shell, during the late-time interactions of some lower-density, slower (and presumably high-metallicity) ejecta with the companion, we may no longer be justified in ignoring the orbital velocity.

An additional important factor in the dynamics of supernovae in binaries is a possible kick imparted to the newly formed neutron star.
This is likely due to a `gravitational tugboat' effect from asymmetry in the ejecta surrounding the neutron star after the core bounce, and
perhaps also high magnetic fields and the asymmetric emission of neutrinos from the proto-neutron star \citep{Kusenko96, Scheck04,
Scheck06, Maruyama11, Wongwathanarat13}.
For \emph{ultra}-stripped supernova progenitors, which have very small ejecta masses, the shock expands very rapidly and the tugboat effect on the neutron star has been shown to be minimal \citep{Suwa15}.
For the range of hydrodynamic simulations in this paper we do not apply any additional kick to the neutron star.

To study this problem hydrodynamically, we simulate a supernova in an orbiting binary self-consistently from just after the moment of core bounce in the supernova.
To this end, we first generate stellar structure models of the binary components using a one-dimensional stellar evolution code, where we strip an evolved massive progenitor of the majority of its envelope.
We then convert these stellar structures to three-dimensional stars in an SPH code, and run simulations from the moment of the SN.
We vary the mass of the primary star as well as the orbital separation independently.
In particular, we are interested in investigating the dependence of the companion's removed mass and impact velocity on the initial orbital separation of the binary.
We describe our numerical method in more detail in the following section.

\section{Method}
Throughout this work we use the Astrophysical Multipurpose Software Environment\footnote{\texttt{www.amusecode.org}} \citep[\AMUSE;][]{AMUSE09, AMUSE13a, AMUSE13b} to perform our simulations.
We first outline the technique used to generate the stellar models in our binary systems (Section \ref{sec:method_stellar_models}).
We then describe the  set up of the initial hydrodynamical models from the stellar structure (Section \ref{sec:method_hydro_setup}).
Finally, we describe the simulation of the SN in the binary, with some discussion on the initial convergence tests that were performed (Section \ref{sec:method_SN}).

\subsection{Stellar models} \label{sec:method_stellar_models}
In order to generate an SPH realisation of the binary, we require a stellar evolution code that can return the internal structure of the star.
Two evolution codes in \AMUSE\ fit this criterion: \MESA~\citep{Paxton11} and \texttt{EVtwin} \citep{Eggleton71, Eggleton06}.
We chose \MESA~to evolve the models to their final stellar structure, motivated by difficulties in previous work in using \texttt{EVtwin} to obtain solutions past the carbon flash in more massive stars \citep{deVries14}.

Due to interactions with the binary companion (and potentially also through stellar winds), much of the mass of the primary star is lost over its lifetime, resulting the helium star progenitor of a Type Ib/c supernova \cite[for some observationally-motivated examples, see][]{Kim15}.
To obtain an estimated lifetime of the progenitor, we first use \texttt{SSE}, which is a fast predictor of stellar properties based on parametrised stellar evolution tracks \citep{Hurley00}.
With the intent of generating $3~\MSun$ and $4~\MSun$ Helium-star progenitors, we begin with a $12.9~\MSun$ and $16.0~\MSun$ zero-age models with metallicity $Z = 0.02$,\footnote{This `canonical' value of the solar metallicity may be an overestimate; see \cite{Asplund09} for a review.} which are predicted by \texttt{SSE} to end with the required helium core masses.

We do not model or speculate on the specific mechanisms of the mass loss from the primary star, but instead apply a constant mass loss (removed from the outer mass shells of the \MESA~structure model) until the final helium star mass is reached.
Because the lifetime in \texttt{SSE} may be an overprediction compared to the actual lifetime reached in \MESA, we apply this mass loss between $80\%$ and $90\%$ of the predicted \texttt{SSE} lifetime so as to not reach the end of the \MESA~evolution before all of the required mass is stripped.
The stellar evolution is then continued until the final lifetime found in \MESA.

For our helium star models, the very final stage of evolution involves a rapid expansion of the remaining, tenuous envelope.
Due to interaction with the close binary companion, this small amount of material in the envelope is in fact expected to be lost from the system; for our progenitors, we use models just prior to this stage, at which the outer radius of the helium star is still compact.
For consistency, we evolve our ($Z = 0.02$, $M_2 = 1~\MSun$) companion star to the same age as that used for the primary star.
In practice, this means the companion is still on the early main sequence, and therefore the effect of this small duration of stellar evolution on the structure and composition of the companion is negligible.

\subsection{Hydrodynamical model set-up} \label{sec:method_hydro_setup}
We model the hydrodynamics of the SN using the SPH code \texttt{Gadget-2} \citep{GADGET}, running in the \AMUSE\ framework.
The SPH formalism has been shown to be effective in three-dimensional simulations of stellar phenomena \citep[for example,][]{Pakmor12}.
One reason is that computational resources are not expended on regions of vacuum or negligible density \citep[for example,~][]{Lai93, Church09}, which constitute a significant fraction of the simulation volume in the current problem.
Modelling a binary in a vacuum is easily handled in SPH, without the need for (low density) background fields in grid codes, which can exhibit artificial shocks from motions of other bodies within this background.

The Lagrangian nature of SPH describes advection naturally, without suffering from complications of numerical diffusion found in Eulerian codes, and we do not have to restrict the simulation to a fixed volume, which is useful in the present problem of a rapidly expanding shell of gas.
A benefit to running the simulation in three dimensions is the absence of any boundary effects, which can produce on-axis artefacts \citep{Marietta00} or preferential wave numbers in the formation of instabilities \citep{Warren13}.
As with all hydrodynamical codes, the SPH method also has its drawbacks, and some of these are discussed further in the context of our convergence studies in Section \ref{sec:method_convergence}.

The stellar models created in \MESA~are converted into SPH particles using the \texttt{star\_to\_sph} routine in \texttt{AMUSE}, in a similar method to that outlined in \cite{deVries14}.
The routine first extracts the one-dimensional hydrostatic structure of the star, represented as a function of mass coordinates, from the data generated by the stellar evolution code.
The SPH particles are initialised in a homogeneous sphere constructed from a face-centred cubic lattice, and the radial positions of the particles are then adjusted so as to match the density profile from the evolution code.\footnote{Randomisation of the angular orientation of the particles has the undesirable effect of the additional shot noise it generates in the initial density distribution; however, the further step of damped relaxation used here will ultimately result in a glass-like configuration.}
The internal energies of the particles are then assigned from the temperature (and mean molecular weight) distribution from the stellar evolution code.
We use equal-mass particles throughout these simulations \cite[unequal-mass particles can cause additional complications such as spurious mixing;][]{Rasio99}.

The primary star is configured with a purely gravitational core particle of $1.4~\MSun$ to model the neutron star.
The softening length $\epsilon$ is chosen to be equal to the smoothing length, such that, due to the compact support of the cubic spline, the smoothing kernel reaches zero at $2.8\, \epsilon$.
This equality is also maintained for the SPH particles to preserve equal resolution of the gravitational and pressure forces.
The zero-kinetic-energy models are relaxed over 2.5 dynamical timescales of the gas using critical damping on the velocities of the particles, where at each step the magnitude of damping is reduced so that in the final step no constraint on the velocity is imposed \citep[for a similar approach, see][]{deVries14}.
This is required due to effects of mapping the one-dimensional stellar structure on to the particle grid, and differences in physics between the codes, such as the value of the adiabatic exponent.

We set up the binary models at different orbital separations, $a$, where the minimum separation is chosen to be greater than the limit of Roche-lobe overflow (RLOF) of the companion star, given by the \cite{Eggleton83} relation,
\begin{equation} \label{eq:rlof}
a_\mathrm{RLOF} = \frac{0.6\,q^{2/3} + \log{\left(1 + q^{1/3}\right)}}{0.49\,q^{2/3}} R_2 ,
\end{equation}
where $R_2$ is the companion radius and $q$ is the binary mass ratio $M_2/M_1$.
Once both stars have been initialised to the SPH code, orbital velocities are determined for a circular orbit at the specified separation and applied to each star.

\subsection{Simulation of the supernova explosion} \label{sec:method_SN}
The SN is initiated using the `thermal bomb' technique \citep{Young07, Hirai14}, which assumes the core bounce has just occurred, at which moment we inject energy into a shell of particles around the neutron star.
As discussed in \cite{Young07}, thermal bomb approaches (along with alternative, piston-driven shocks) are not intended to embody the physical mechanism that drives the SN.
Indeed, the actual processes by which the energy gain occurs near the proto-neutron star are still not fully understood, though recent observations and insights from three-dimensional simulations have shed some light on the role of instabilities, asymmetries and jets in driving this process \citep{Janka12, Bruenn13, Hanke13, Lopez13, Milisavljevic13, Couch13, Couch15}.

The boundary of energy injection is specified by radius (which is equivalent to a fixed enclosed mass) instead of particle number.
This allows scaling of the problem over a range of SPH particle numbers while keeping fixed the mass fraction that receives the SN energy.
The total thermal energy (a canonical $10^{51}~\mathrm{erg} \equiv 1~\mathrm{foe}$\footnote{More recently, the `Bethe' ($\mathrm{B}$) has been proposed as an equivalent unit in honour of Hans Bethe's work on supernovae \citep{Weinberg06, Woosley07}.}) is distributed evenly amongst these $N_\mathrm{SN}$ particles, so that the specific internal energy per particle is increased by $\left(1~\mathrm{~foe}\right)/\left(N_\mathrm{SN} \, m_\mathrm{SPH}\right)$.

We found that a careful investigation of the effect of the injection radius was necessary.
Too small a radius (and therefore $N_\mathrm{SN}$) results in large, uncontrolled asymmetries in the shock front that grow from intrinsic small-scale asymmetries in the initial particle distribution.
On the other hand, too large a radius results in the internal energy of the SN being distributed amongst a large number of particles, lowering the specific internal energy and therefore reducing the overall temperature in the region and weakening the shock.
We found that, for the helium star models used here, injecting the SN energy into a region $R_\mathrm{SN} \lesssim 0.05 \RSun$ generates a sufficiently spherical shock while still keeping $N_\mathrm{SN}$ small as possible.

Two of the parameters we wish to predict, calibrated from our simulations, are the final velocities (formally, at infinity) of the runaway components of distributed binaries.
For masses $m$ relative to the neutron star mass (i.e. $m \equiv M/M_\mathrm{NS}$), TT98 predict these values in terms of the following initial parameters
\begin{itemize}
\item $a$: the pre-SN binary orbital separation
\item $v$: the pre-SN relative orbital velocities
\item $w$: the magnitude of the kick applied to the NS
\item $\theta$ and $\phi$: the spherical polar angles of the NS kick vector with respect to the `$x$'--axis aligned along the NS orbital velocity vector at the moment of the kick
\item $v_\mathrm{im}$: the change in velocity of the companion due to the impact of the SN shell
\item $v_\mathrm{ej}$: the velocity of the ejecta shell
\item $m_2$, $m_{2\mathrm{f}}$ and $m_{\mathrm{shell}}$: the initial mass of the companion, the final mass of the companion after mass loss, and the mass of material in the shell, respectively (all relative to the neutron star mass)
\end{itemize}

In the original work of \cite{Wheeler75}, during the supernova shell passage over the companion star, the effect of mass stripping is parametrised by the fraction of companion radius $x = R/R_2$ as a function of angle around the star.
Above some critical fraction of the companion radius, $x_\mathrm{crit}$, a fraction $F_\mathrm{strip}$ of the mass is stripped by the shell impact, and below it a fraction $F_\mathrm{ablate}$ of mass is ablated.

In this framework, the magnitude of the impact velocity $v_\mathrm{im}$ is theoretically predicted to be
\begin{align} \label{eq:impact_velocity}
v_\mathrm{im} &= \eta \frac{\Delta p}{M_2\left(1 - F^*\right)} \nonumber \\
              &= \eta v_\mathrm{ej} \left(\frac{R_2}{2a}\right)^2 \frac{M_\mathrm{ej}}{M_2} x_\mathrm{crit}^2 \frac{1 + \ln{\left(2 v_\mathrm{ej}/v_\mathrm{esc}\right)}}{1 - F^*} .
\end{align}
Here, we use the prediction by \cite{Wheeler75} in the form adopted by \cite{Tauris15}, which applies the substitution $\left(F_\mathrm{strip} + F_\mathrm{ablate}\right) = F \rightarrow F^* = \left(F_\mathrm{strip} + F_\mathrm{ablate}\right)^\alpha$ for some $\alpha$, as well as a free parameter $\eta$ to account for the fact that this tends to over-predict the value of $v_\mathrm{im}$.
Effectively, $\eta$ represents the final change in momentum of the companion as a fraction of the incident momentum in the shell.
The fraction of mass stripped and ablated from the companion, $F_\mathrm{strip}$ and $F_\mathrm{ablate}$, is calculated in \cite{Wheeler75} using a polytropic star of index $n = 3$.

As noted in \cite{Wheeler75}, corrections must be applied to this formula as it neglects the presence of a rarefaction wave back through the ejecta, geometrical effects of curvature in the shell (more important for small $a$), inhomogeneities in the ejecta and radiative losses behind the shock.
Further phenomena can also modify the effective impact velocity, such as the deformation of the companion by the shock passage (altering its cross-sectional area), the formation of a bow shock in the ejecta (during which time the flow deflects around the companion star), and shock convergence on the far side of the star (causing the asymmetric emission of material from this side of the star).

One key element in the predictions of TT98 is the estimate of the mass stripped from the companion.
This is based on early work from \cite{Wheeler75} and simulations of a planar slab of material hitting a star \cite{Fryxell81}, which have a low resolution by today's standards.
As noted above, results based on plane-parallel ejecta profiles need correcting for the fact that the shell in reality has some curvature.
Higher resolution simulations such as the present one provide a test of these early estimates, which are one of the sources of uncertainty in the results of TT98.

The stripping of mass in our simulations is measured by calculating the specific energy for each particle,
\begin{equation} \label{eq:bound_mass}
 e_\mathrm{tot} = e_\mathrm{kin} + e_\mathrm{therm} + e_\mathrm{pot} = \frac{1}{2}v^2 + u - \phi,
\end{equation}
which is negative for bound particles.
The amount of bound mass in the companion is time-dependent over the course of the simulation due to energy exchange between particles.
The stabilisation of mass bound in the companion determines the end of our simulation, which in practice occurs within 10 dynamical timescales of the companion star ($\sim 2 \times 10^4~\mathrm{s}$).

\subsubsection{Convergence test} \label{sec:method_convergence}
One feature of SPH that demands caution is that the resolution is dependent on the local density, and therefore the method loses resolution in the lower-density, uppermost layers of the stars in our simulations.
In the current problem, the mass stripped by the secondary is from these same layers.
Therefore, a good test for the resolution of the simulations is to look for convergence in the quantity of mass stripped from the companion.

During the SN, Richtmeyer--Meshkov (the impulsive form of Rayleigh--Taylor) instabilities (RMI) are expected to be present, which have been found to appear once reverse shocks form at the interfaces between discontinuities in the density gradient \citep{Kane99}.
Such discontinuities are present in Type Ib/c progenitors at the interface between the carbon-oxygen boundary in the core and, if any substantial fraction of hydrogen remains in the envelope, also at the helium-hydrogen boundary.
However, these discontinuities tend to be smoothed during the conversion from the 1D stellar model and subsequent relaxation of the SPH particles.
Proper treatment of the RMI requires a prescription of artificial conductivity that is not included in the current SPH codes in AMUSE.
This instability is expected to be a significant factor in the mixing of stellar material early in the evolution of supernova remnants (SNRs), and so any evaluation of the fate of the composition of the SN ejecta must take this into account.

Unless the growth of RMI is explicitly seeded by some structure at the density interface, these instabilities will grow from perturbations at the numerical level of the simulation and may not, in such cases, grow substantially \citep{Kane99}.
Therefore, there is the potential for instabilities to be dependent on numerical effects such as the resolution of the simulation.
Additionally, during the stripping of mass from the companion star, the initial deceleration of the shell impacting the companion may be Rayleigh-Taylor unstable, but also that the subsequent flow of the shell over the surface of the star may induce some shearing (Kelvin--Helmholtz) instabilities (KHI).

Due to the smoothing of discontinuities after relaxation of the SPH models, a lack of artificial conductivity\footnote{This smooths thermal energy discontinuities and is used in capturing the vortices seen in KHI. However, there has been some debate on the causes of KHI suppression in SPH; see, for example, the discussion in \cite{Gabbasov14}.} in \texttt{Gadget-2} and the only perturbations being from noise in our SPH distribution, we expect that instabilities will not be fully captured in our simulations.
Since the overall capturing of instabilities is reduced, so too should be the effect of instabilities on our results.

\begin{figure}
\includegraphics[width=0.95\columnwidth]{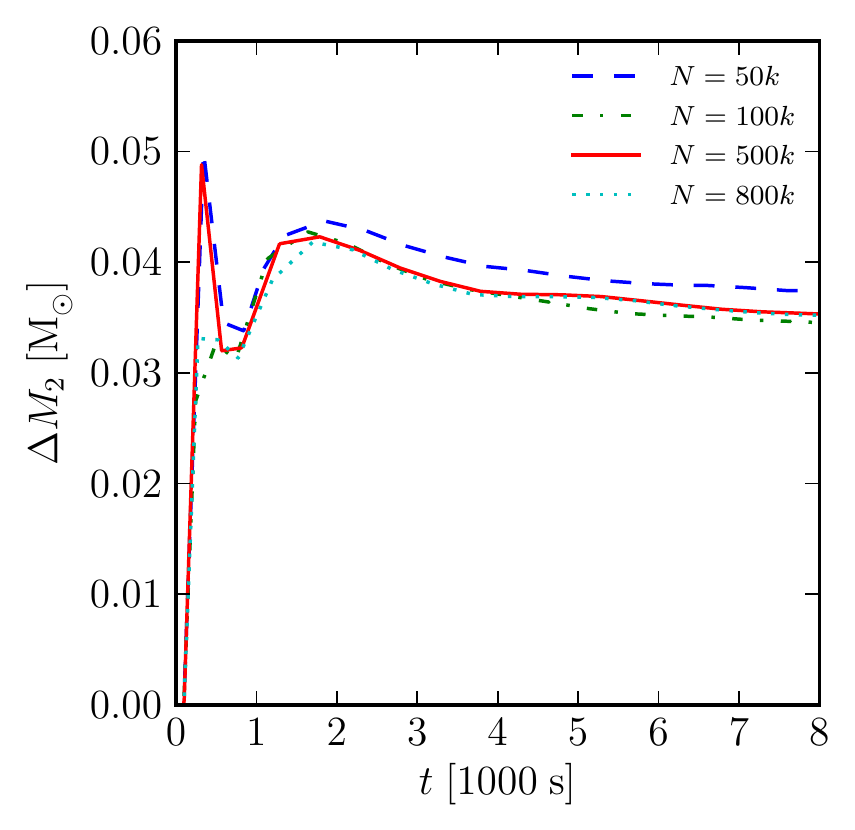}
\caption{Results for a convergence study using the amount of mass stripped from the companion.
The primary star mass was $3~\MSun$ and the orbital separation was $4~\RSun$ in all cases.
The number of SPH particles used in each run is given in the legend.\label{fig:convergence}}
\end{figure}

Figure \ref{fig:convergence} shows a test of varying the SPH particle number, $N$, based on the amount of mass lost from the companion star (evaluated using equation \ref{eq:bound_mass}).
For low $N$, there is noticeable noise in the bound mass determination over time, but for $N \geq 10^5$ particles, this is no longer appreciable.
As shown in Figure~\ref{fig:convergence}, we did not find any substantial difference in the results increasing $N$ from $5 \times 10^5$ to $8 \times 10^5$.
Accounting for this, as well as available computational resources, our simulations were run with $5 \times 10^5$ particles.

\section{Results}
After reviewing the initial conditions used for our simulations, we examine the early stages of the SN (Section \ref{sec:results_breakout}).
We then investigate the magnitude of mass lost from the companion as a function of the orbital separation (Section \ref{sec:results_mass_loss}), as well as the velocity imparted to the companion and the fraction of imparted momentum compared to the incident shell (Section \ref{sec:results_velocities}).
Next, we examine the newly formed SNR for asymmetries in morphology and metallicity (Section \ref{sec:results_snr}).
Finally, we consider the subsequent evolution of a star altered by a SN shell impact (Section \ref{sec:results_evolution}).

Table \ref{tab:ic} shows the initial conditions used in our simulations.
The choice of primary and companion masses is motivated by the binary parameters used in TT98 and \cite{Tauris15}, while the minimum orbital separations are chosen to be outside the RLOF value (equation \ref{eq:rlof}).
The final two columns show the effects on the companion due to the shell impact, discussed in more detail in the remainder of this section.

\begin{table}
\begin{tabular}{ccccc}
$M_1$ $(\MSun)$ & $M_\mathrm{ej}$ $(\MSun)$ & $a$ $(\RSun)$ & $\Delta M_2$ $(\MSun)$ & $v_\mathrm{im} (\mathrm{km\,s}^{-1})$ \\
\hline
4.0 & 2.6 & 4.5 & 0.021 & 78 \\
4.0 & 2.6 & 5.5 & 0.013 & 57 \\
4.0 & 2.6 & 6.5 & 0.0096 & 47 \\
3.0 & 1.6 & 4.0 & 0.037 & 83 \\
3.0 & 1.6 & 5.0 & 0.020 & 60 \\
3.0 & 1.6 & 6.0 & 0.013 & 48 \\
\end{tabular}
\caption{Simulation initial conditions and main results.
The first three columns indicate the initial conditions, where $M_1$ is the mass of the primary (helium star) and before the SN, $M_\mathrm{ej}$ is the total ejecta mass, and $a$ is the initial orbital separation.
The last two columns are the amount of mass stripped from the companion star and the (magnitude of the) impact velocity.
\label{tab:ic}}
\end{table}

\subsection{Shock breakout} \label{sec:results_breakout}
By approximately $20~\mathrm{s}$ after the SN is initiated, the forward shock has broken out of the surface of the helium star, during which time the fraction of SPH particles bound to the $1.4~\MSun$ neutron star drops smoothly to almost zero.
We find at late times that there is some fall-back of a small amount of material, which remains bound to the neutron star.
As we do not model here the complexities of the magnetic field of the new neutron star or any form of pulsar wind, it is possible that other mechanisms later expel some or all of the residual bound gas.

\begin{figure}
\includegraphics[width=0.95\columnwidth]{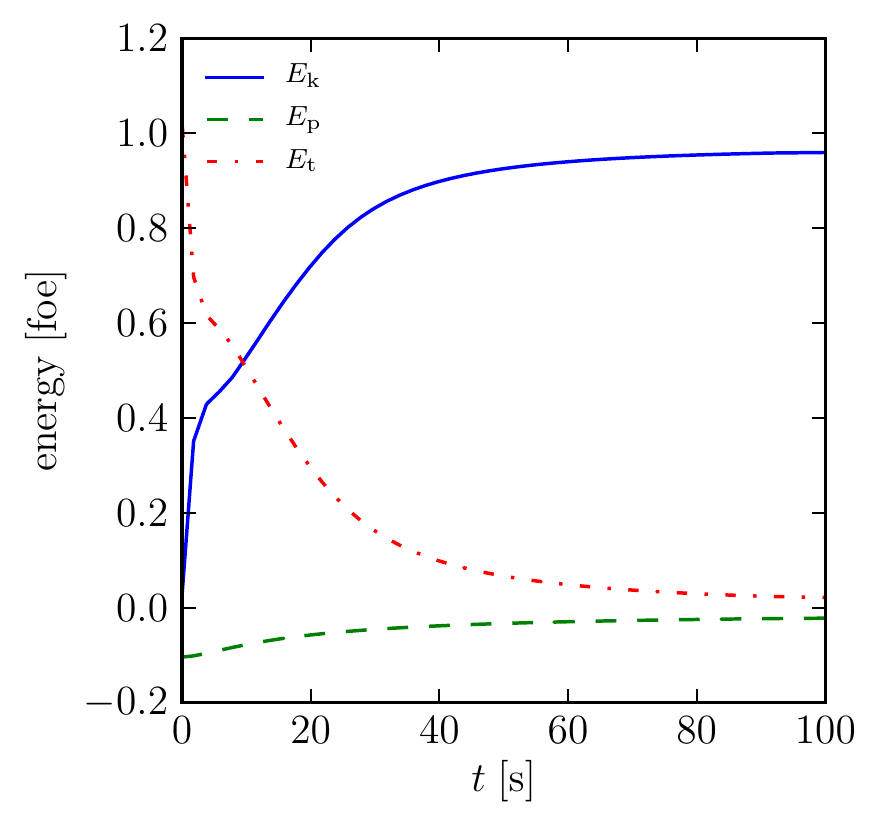}
\caption{Distribution of total energy in the gas, in units of $\mathrm{foe}$ ($10^{51}~\mathrm{ergs}$), as a function of time following the supernova event.
The energy is broken down into kinetic ($E_\mathrm{k}$), potential ($E_\mathrm{p}$) and internal (thermal) ($E_\mathrm{t}$).
This example corresponds to a primary of mass $4~\MSun$ and an orbital separation of $4.5~\RSun$
\label{fig:energies}}
\end{figure}

In Figure \ref{fig:energies}, it can be seen that there is a rapid conversion of energy from internal (thermal) energy from the moment of explosion to kinetic and potential energy as the shock passes through the star and the subsequent shell expands.
By approximately $100~\mathrm{s}$ following the SN explosion, very little of the original thermal energy  remains in the gas as it has been almost entirely converted into kinetic energy in the expanding shell.

\begin{figure*}
\includegraphics[width=1.95\columnwidth]{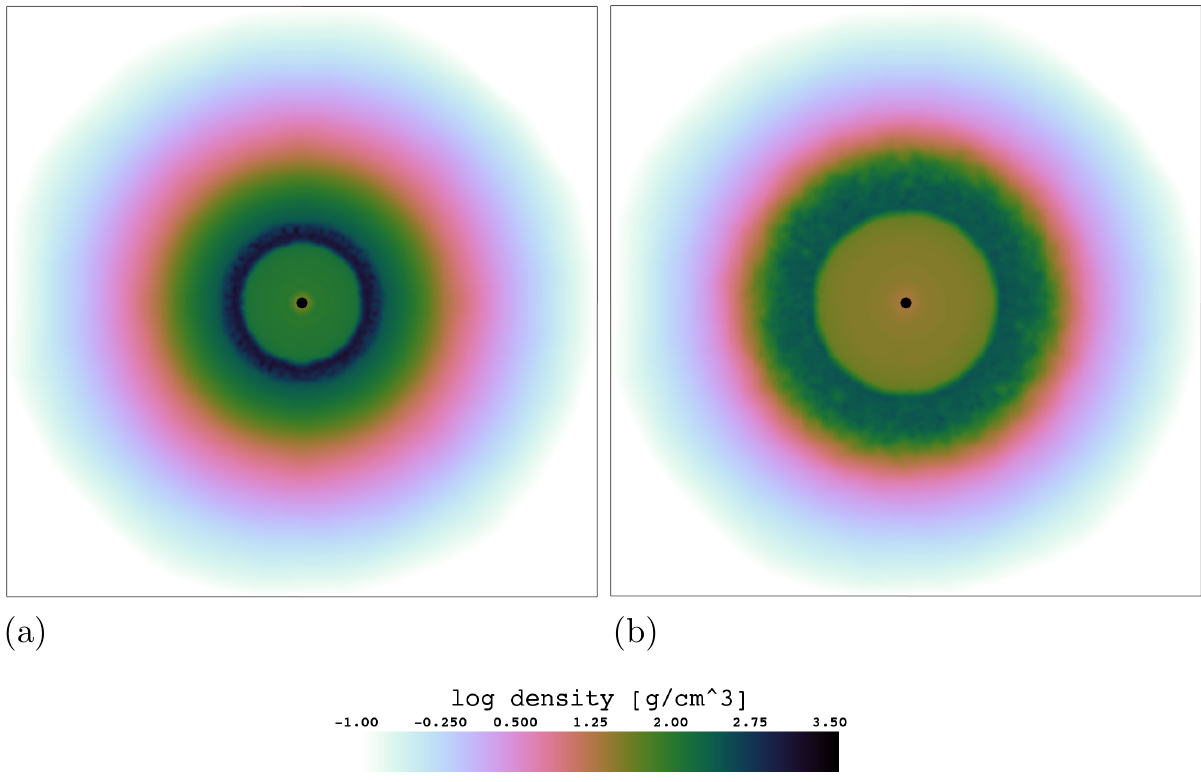}
\caption{A density slice through the $x$-$y$ (initial orbital) plane during the SN explosion in a $3.0~\MSun$ primary at (a) $4~\mathrm{s}$ and (b) $24~\mathrm{s}$ after the initiation of the SN.
The boxes are $2~\RSun$ along each side.
The black point at the center of the star is the (gravitational) $1.4~\MSun$ NS particle.
\label{fig:SN_snapshots}}
\end{figure*}

Figure \ref{fig:SN_snapshots} shows the changes in density through the $3~\MSun$ helium star from shortly after core bounce until around the time the forward shock reaches the outer layers of the star.
A lower density cavity with a very shallow gradient is seen to lag behind the expanding ejecta shell, which has a steepening density gradient as a function of radius towards the leading edge of the ejecta.
This variation in gradient is maintained over time, although the overall magnitude of the density drops during the expansion.
We investigate the density and velocity distributions within the ejecta in more detail in Section \ref{sec:results_velocities}.

\subsection{Impact and mass loss from the companion} \label{sec:results_mass_loss}
The passing shell first strips material from the outer layers of the companion.
The compression of the companion along the direction of motion of the shock causes heating of the stellar material, which results in a subsequent mass loss through ablation.
This ablation of material is found to be a slower form of mass loss than the initial stripping phase.

\begin{figure*}
\includegraphics[width=1.95\columnwidth]{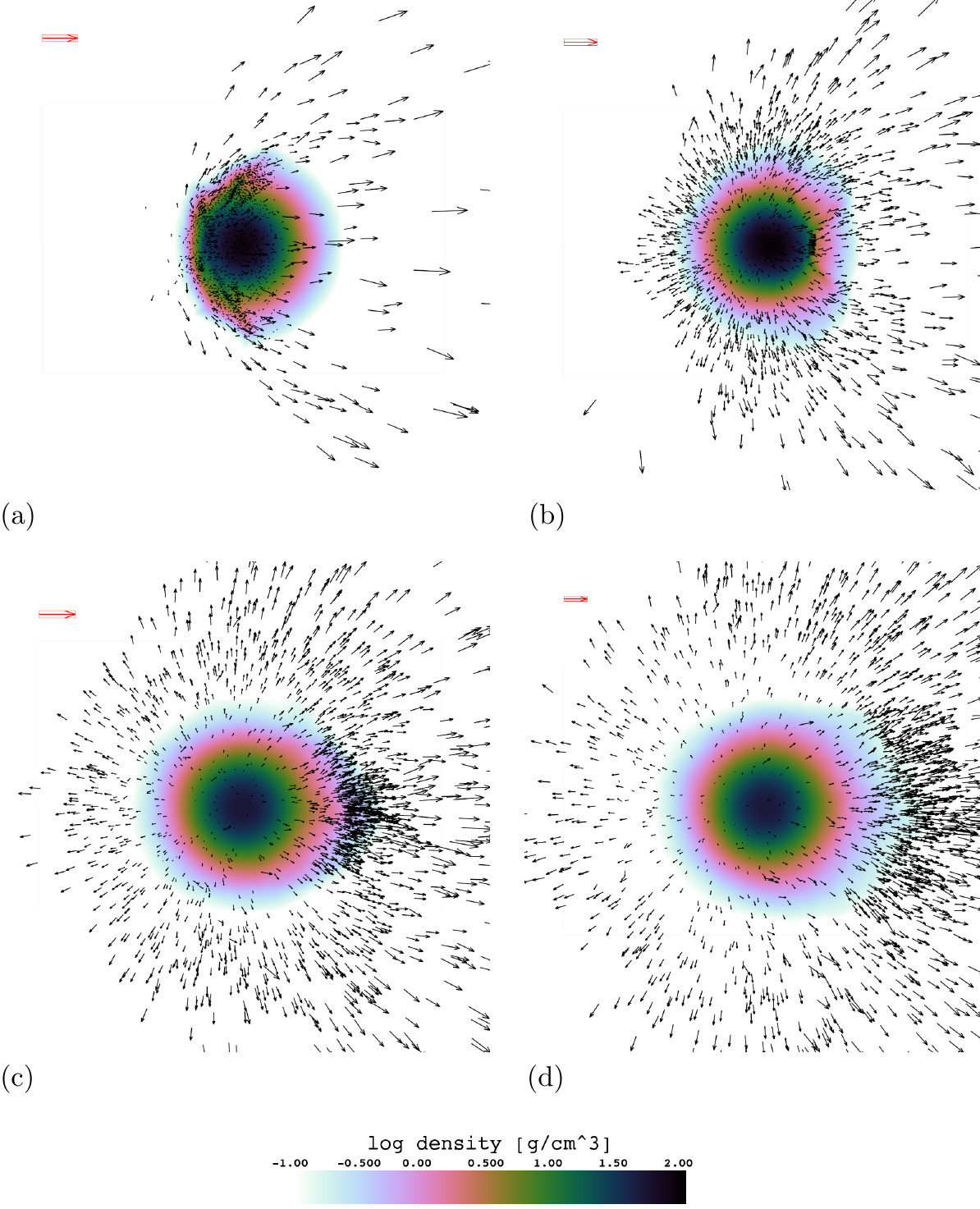}
\caption{A slice through the $x$-$y$ (initial orbital) plane during the passage of the SN shock through a $1.0~\MSun$ companion star after a $1~\mathrm{foe}$ SN in a primary star of $3.0~\MSun$ at a distance of $4~\MSun$.
The snapshots correspond to times of (a)  $433~\mathrm{s}$, (b) $1028~\mathrm{s}$, (c) $1628~\mathrm{s}$ and (d) $2028~\mathrm{s}$.
The shock enters the companion star from the left.
The black vectors show the magnitude of the velocity projected onto the orbital plane (and with the orbital velocity of the companion subtracted) for a small random sample of all the particles removed from the companion.
In each case, a reference vector (red, boxed) is given in the upper left corner; these correspond to (a) $10^{4}~\mathrm{km\,s}^{-1}$, (b) $3 \times 10^{3}~\mathrm{km\,s}^{-1}$, (c) $2 \times 10^{3}~\mathrm{km\,s}^{-1}$ and (d) $10^{3}~\mathrm{km\,s}^{-1}$. 
\label{fig:companion_snapshots}}
\end{figure*}

The passage of the shock through the companion can be seen in the density slices of Figure~\ref{fig:companion_snapshots}.
Aside from the mass stripping from the sides of the star as predicted in \cite{Wheeler75}, there is also mass loss from the far side after the shock has passed through the star.
The black vectors in this figure show the velocities for a random sample of all the SPH particles that were originally in the companion which have subsequently become unbound.
These vectors have had the orbital velocity vector of the companion subtracted, and they are then projected onto the orbital plane.
Because each SPH particle has the same mass, these vectors also indicate the relative momentum of the unbound particles.

Panel (b) of Figure~\ref{fig:companion_snapshots} shows that, once the shock passes through the centre of the companion, it converges at the far the side of the star as it accelerates down the density gradient \citep[similar shock convergence is seen around other spherically symmetric density gradients, such as in][]{Rimoldi15}.
This increases the local pressure on this axis, resulting in expulsion of material from the far side of the star and can counter the effect of the outward kick imparted by the incident shell of material \citep[see also][]{Marietta00}.

In the last panel of Figure~\ref{fig:companion_snapshots}, the central density of the companion has dropped and it has noticeably expanded from the shock heating.
During this later period (final three panels) ablation occurs for material whose specific internal energy is greater than the square of the sound speed.
Due to the shock heating, the companion is `puffed-up' like a pre-main-sequence star, and its luminosity is expected to increase temporarily as it reverts to thermal equilibrium \citep{Marietta00}.
Finally, we find a quadrupole oscillation of the companion that is induced by the distortion from compression due to the shock
This ringing subsides after about one dynamical timescale of the companion star.

\begin{figure}
\includegraphics[width=0.95\columnwidth]{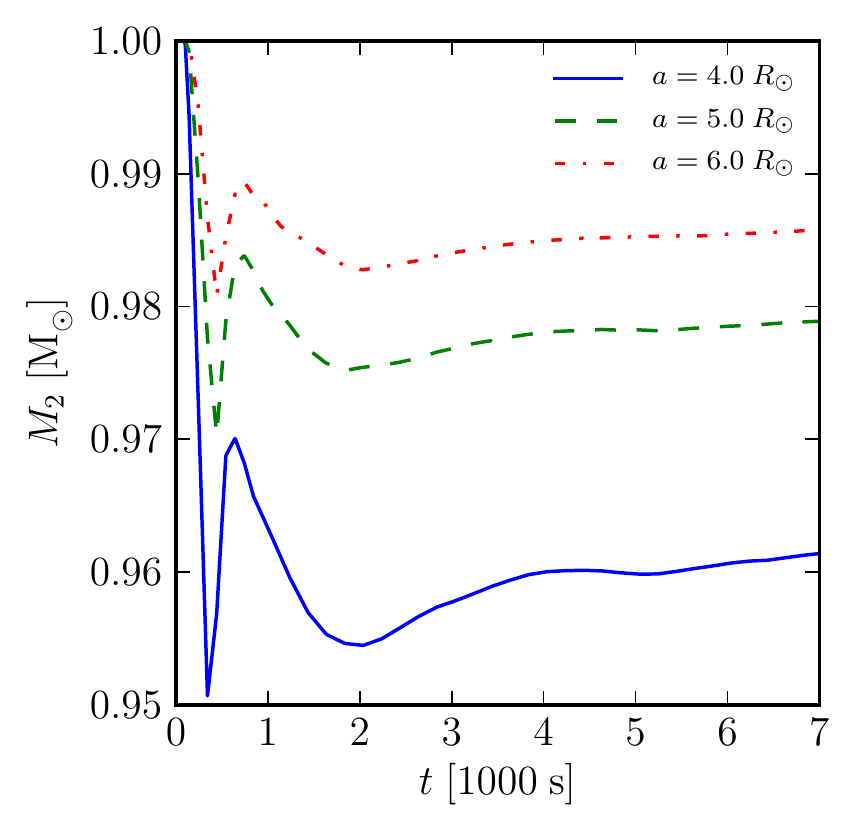}
\caption{Mass bound to the companion using equation (\ref{eq:bound_mass}) as a function of time since the SN explosion for a primary helium star of $3.0 \MSun$ and a range of orbital separations.\label{fig:bound_mass}}
\end{figure}

Figure \ref{fig:bound_mass} shows an example of the variation in companion mass due to the shell impact.
The stripping of mass by the passing shell causes a rapid mass loss in the initial phase, followed by a more gradual mass loss due to the later ablation of shock-heated material.
The proportion of mass lost drops rapidly even by moderate orbital separations.

\begin{figure}
\includegraphics[width=0.95\columnwidth]{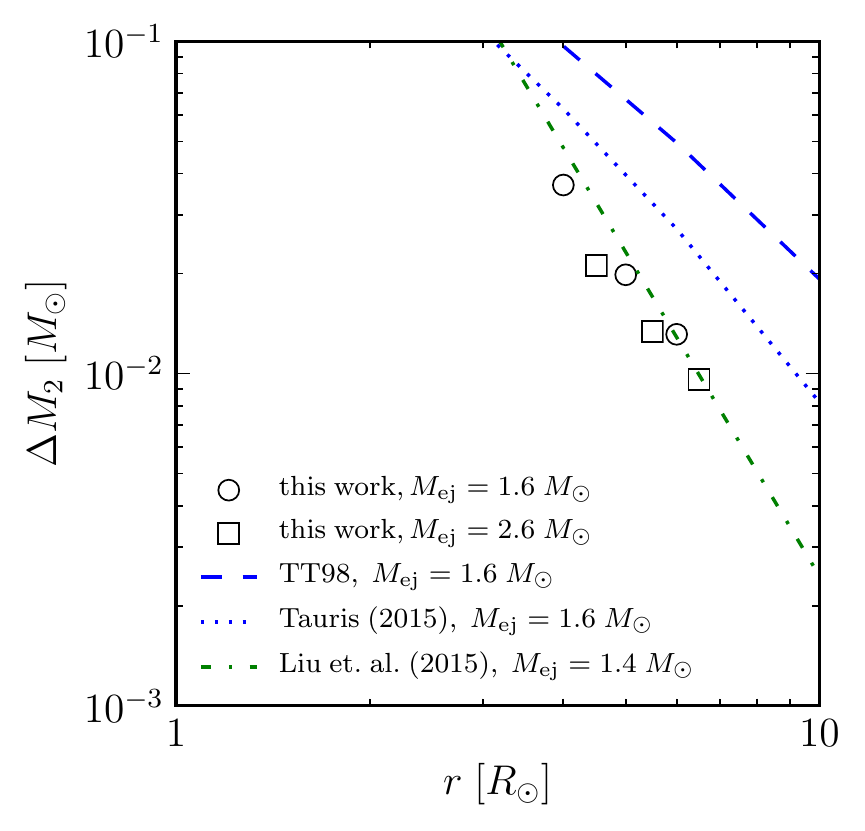}
\caption{Final mass lost from the companion as a function of orbital separation. Circles show results with a primary star of $3~\MSun$ and squares show results for a primary star of  $4~\MSun$.
The secondary is $1~\MSun$ in each case.
The comparison curves are from the theoretical predictions of WLK75 as adapted by TT98 and Tauris (2015) (rescaled to our initial conditions), as well as the simulation results of Liu (2015) for a $0.9~\MSun$ companion star.
Note that the comparison with Liu (2015) is not fully equivalent, as both the ejecta mass and companion mass (and therefore radius) are slightly different.
\label{fig:dm_vs_a}}
\end{figure}

In Figure~\ref{fig:dm_vs_a}, we show the amount of mass lost from the companion (as a fraction of its initial mass) as a function of orbital separation.
The lost mass is found by subtracting the final bound mass at the last snapshot of each simulation (which occurs at $2 \times 10^4~\mathrm{s}$) from the initial mass.
We use the last time possible from the simulation as the final mass takes much of the total simulation time to reach its steady-state value.
The dashed line shows the prediction from TT98, the dotted line shows the fit obtained from Type~Ia simulations compiled by \cite{Tauris15}, and the dot-dashed green line is from the new work of \cite{Liu15}.
A least-squares regression gives a fit to our data of $1.3 \left(R/\RSun\right)^{-2.6}$ for the $M_\mathrm{ej} = 1.6~\MSun$ data and $0.58 \left(R/\RSun\right)^{-2.2}$ for the $M_\mathrm{ej} = 2.6~\MSun$ data.
We find comparable values of the lost mass, in particular the steeper power-law gradients, to those seen in \cite{Liu15}.

There are some differences between our initial conditions and those from the previous work shown in Figure~\ref{fig:dm_vs_a}.
Compared with the calibration from Type~Ia simulations, our explosion energies and ejecta masses are both slightly different.
Additionally, the $1~\MSun$ companion radius, $R_2$, shrinks slightly after relaxation of the SPH models compared with the canonical $1~\RSun$.
The predictions in TT98 and \cite{Tauris15} depend on these quantities in particular within the geometric parameter $\Psi$, used originally by \cite{Wheeler75}, defined as
\begin{equation}
 \Psi = \left(\frac{R_2}{2 a} \right)^2 \left( \frac{m_\mathrm{shell}}{m_2} \right) \left( \frac{v_\mathrm{ej}}{v_\mathrm{esc}} - 1 \right) .
\end{equation}
This parameter is used in the determination of $x_\mathrm{crit}$ as well as $F_\mathrm{strip}$ and $F_\mathrm{ablate}$ in \cite{Wheeler75} using tabulated data for an $n = 3$ polytrope.
For our comparisons, we adjust these quantities (and therefore $\Psi$) in the TT98 and \cite{Tauris15} estimates to match the initial conditions of our simulations.
Furthermore, the simulations in \cite{Liu15} also use a slightly different companion mass and ejecta mass, and so their results are not completely equivalent to ours.

\subsection{Momentum transfer and the velocity of the companion} \label{sec:results_velocities}
When the orbital separation is very small, the impact of the ejecta causes not only a significant loss of mass from the companion star but also a large change in velocity.
The largest change in velocity of the companion occurs during the transfer of momentum from the shell in the initial impact.
However, as the end of the shell passes over the far side of the companion, there is an overpressure acting on this side of the star when the shock converges on this axis.
This causes the companion to receive a slight change in momentum in the direction opposite to the shell motion \cite[which has been suggested in other simulations such as][]{Fryxell81, Marietta00}.
In the theory of TT98, $\vect{v_\mathrm{im}}$ is defined to be an \emph{effective} velocity that not only accounts for the momentum imparted to the companion by the passage of the shell but also the subsequent change in momentum due to (potentially asymmetric) mass loss.

We found that measuring the velocity of the companion with respect to the compact remnant is complicated by the difficulty to define the baryonic centres of the binary system with the ejecta that had not yet left the binary system, the oscillatory behaviour of the companion star as a result of the shell impact, and the Brownian motion of the neutron star due to the shot-noise of the limited resolution it its vicinity.

As an alternative technique, we set up a co-rotating frame of reference that matches the original circular orbital motion.
Before the SN and up until the shell impact, there is no component of velocity of the companion perpendicular to this direction of motion.
However, during and after the impact, the companion (as well as the mass unbound from it) gains a component of velocity, and therefore momentum, in the radial direction with respect to this frame.
We use this to measure the momentum delivered to the companion and the material removed from the companion, as well as the effective kick velocities.

\begin{figure*}
\includegraphics[width=1.95\columnwidth]{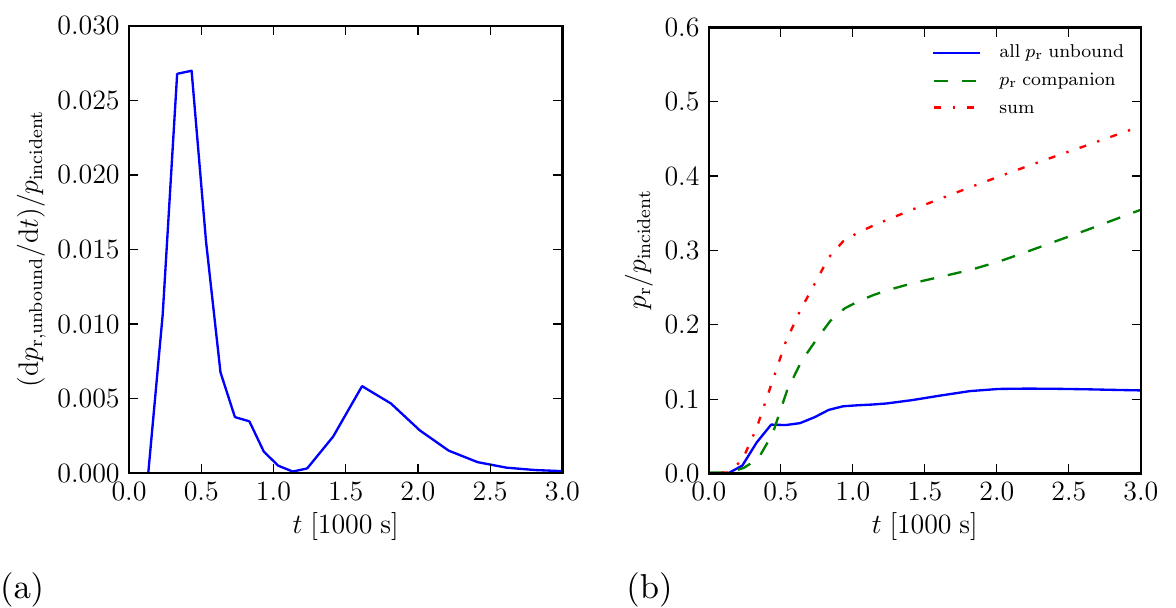}
\caption{Components of momenta in the radial direction of (a) newly unbound material from the companion and total momenta in the companion and  (b) total unbound mass, bound mass in the companion and the sum.
Values are shown as a fraction of the total incident momentum calculated for the cross-section of shell material that impacts the companion.
The example shown is for a $3~\MSun$ primary and an orbital separation of $6~\RSun$
\label{fig:momenta}}
\end{figure*}

Figure \ref{fig:momenta} shows the component of momentum in the radial direction for material unbound from the companion that was not unbound in the previous time step.
This allows us to see clearly the burst out of the back of the star; we measure the impact velocity just after this peak.
The right panel shows the breakdown of momenta in the radial direction for unbound and bound material originally from the companion (and their sum).
This gives an alternative indication of $\eta$, where we see that although less than half of the total incident momentum in the shell is delivered to this material in total, only $\lesssim 30\%$ of the momentum is delivered to the (bound material of the) companion star.

\begin{figure}
\includegraphics[width=0.9\columnwidth]{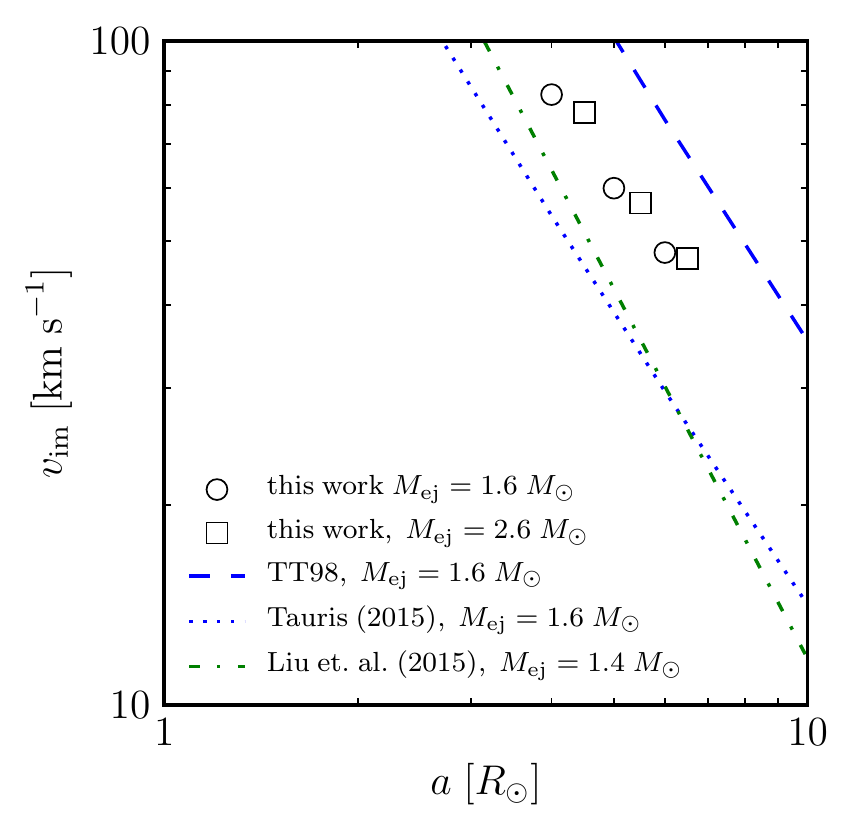}
\caption{Magnitude of the impact velocity, $v_\mathrm{im}$, imparted to the companion star as a function of orbital separation $a$.
Markers and line styles correspond to those used in Figure~\ref{fig:dm_vs_a}.
Again, as for Figure~\ref{fig:dm_vs_a}, the comparison with Liu (2015) is not fully equivalent due to the slightly different ejecta mass and companion parameters.
\label{fig:v_vs_a}}
\end{figure}

Our final impact velocity magnitudes are shown in Figure~\ref{fig:v_vs_a}.
A least-squares regression gives a fit to these data of $556 \left(R/\RSun\right)^{-1.4}$ for the $M_\mathrm{ej} = 1.6~\MSun$ data and $652 \left(R/\RSun\right)^{-1.4}$ for the $M_\mathrm{ej} = 2.6~\MSun$ data.
The velocities for both ejecta mass conditions follow a similar gradient to earlier work presented in TT98 and \cite{Tauris15}, although it is not quite as steep as the $-1.9$ power-law of \citep{Liu15}.
The overall scaling differs from previous work, however.
The early estimate from TT98 of the impact velocities used a value of $\eta = 0.5$ (see equation \ref{eq:impact_velocity}), whereas fits in \cite{Tauris15} and \cite{Liu15} sit closer to $\eta = 0.2$.
Our impact velocities lie in between these values, and we consider one possible cause in the Section~\ref{sec:compare_ejecta}.

Finally, we also consider the effect of drag from the remaining material on the companion velocity, noting that there is still a non-negligible density of gas interior to the ejecta shell.
For a conservative estimate of this drag force from the innermost ejecta, we neglect any outward velocity of this gas, and take a density of $10^{-3}~\mathrm{g\,cm}^3$ in this material.
With these values, the drag force on the companion will be
\begin{equation}
 F_\mathrm{drag} = \frac{1}{2} \rho v_2^2 C_\mathrm{drag} A_2 \approx 2 \times 10^{28}~\mathrm{N},
\end{equation}
for $v_2 = 300~\mathrm{km\,s}^{-1}$, and where we approximate the drag coefficient of the star with a solid sphere value of $C_\mathrm{drag} \approx 0.5$.
For a companion mass of $M_2 = 1~\MSun$ the acceleration associated with this drag is therefore only $10^{-5}~\mathrm{km\,s^{-2}}$.
Although small, drag induced by the lower-velocity ejecta may alter the final effective impact velocity when integrated over a long timescale.

\subsubsection{Ejecta profiles} \label{sec:compare_ejecta}
We investigate in more detail our ejecta profiles as a potential cause of the discrepancy between our impact velocities and those of \cite{Liu15}.
Previous work, such as that of \cite{Liu15}, has often initialised the ejecta with the assumption that it is in a homologous expansion by the time it impacts the companion, so that, for a given $t$, $v \propto R$.
The density and velocity profiles in this ejecta are constructed from broken power-law fits to analytic treatments of the shock through the progenitor.
These treatments have, in particular, been based on the polytropic envelopes (or one-dimensional structure models) of supergiant stars, and the power-law fits are to the (small and large $R$) asymptotic limits of a varying density gradient in the ejecta \citep{Chevalier89, Matzner99}.

\begin{figure*}
\includegraphics[width=1.95\columnwidth]{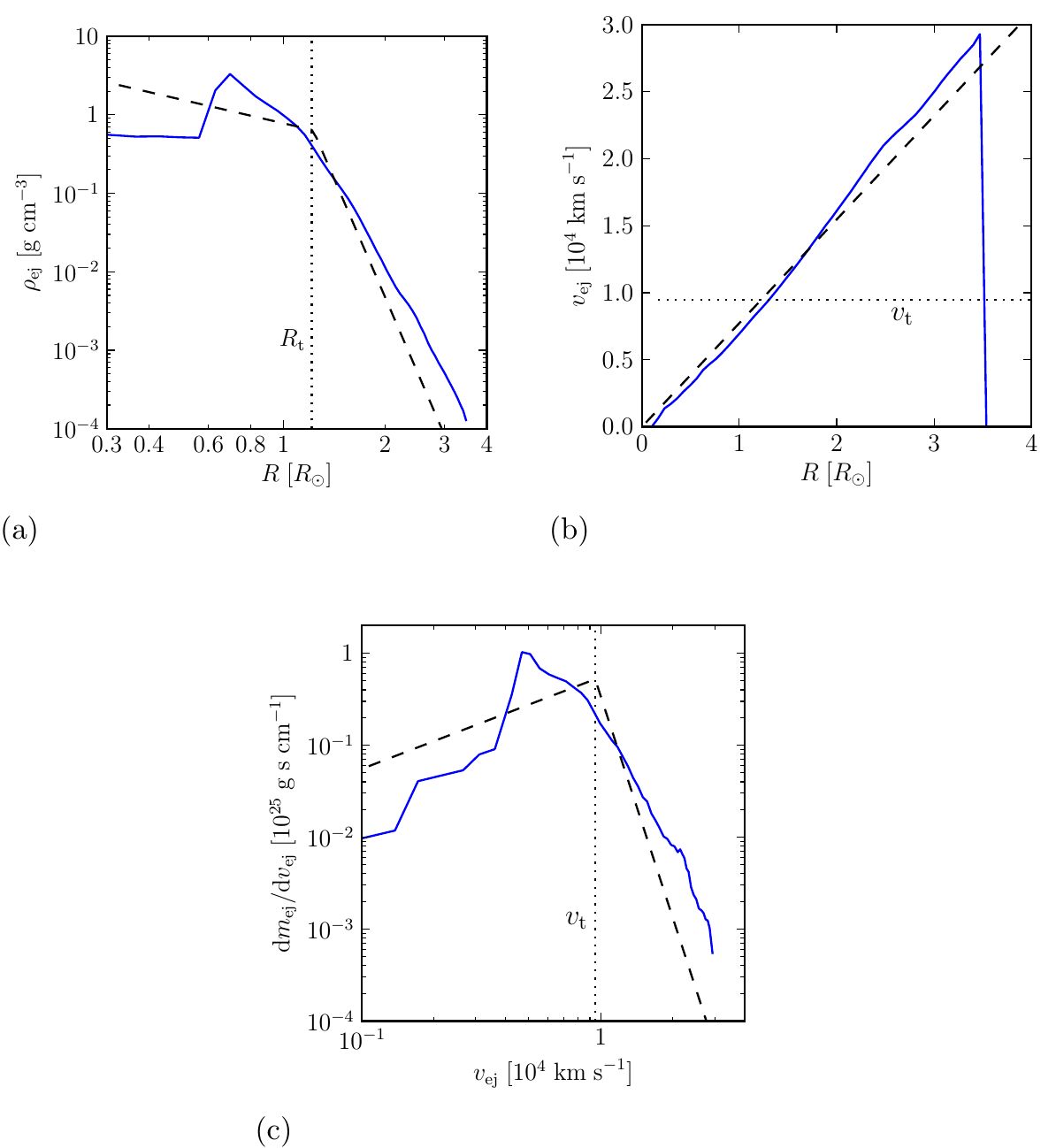}
\caption{Blue lines show the profiles of (a) density, (b) velocity and (c) the mass distribution of velocity within the ejecta for our simulations (binned in radial shells form the center of mass of the ejecta).
Black dashed lines show the power-law profiles used in Liu (2015).
Both cases were calculated for a time of $90~\mathrm{s}$ after the supernova.
The transition velocity (and radius at which this occurs) in the Liu (2015) profiles are given as dotted lines.
\label{fig:compare_ejecta}}
\end{figure*}

In Figure~\ref{fig:compare_ejecta}, we show the variation of ejecta density and velocity as a function of radius from the center of mass (by averaging the SPH particles over concentric shells) and compare with an analytic function from the equations used in \cite{Liu15}.
We also show, in the bottom panel of Figure~\ref{fig:compare_ejecta}, the distribution of velocity over mass in the ejecta.
It is clear from Figure~\ref{fig:compare_ejecta} that in the ejecta from our helium star models we have a shallower density gradient through much of the outer regions compared with the power-law profiles.
In this outer ejecta, the velocity and density are also higher in our models.
As the kick velocity has a strong dependence on this high-velocity ejecta \citep{Liu15}, this can explain increased impact velocities seen in our simulations.

Finally, we note that we examined the ejecta for large-scale asymmetries by determining the shell-averaged radial profiles of density and velocity in hemispheres corresponding to the directions toward and away from the companion star.
We found that the values in either direction agreed to within a few per cent, and therefore do not produce a discernible difference on the logarithmic plots in Figure~\ref{fig:compare_ejecta}.

\subsection{Properties of the larger-scale supernova remnant} \label{sec:results_snr}
The amount of accretion on the companion has previously been shown to decrease with increasing shell velocity \citep{Fryxell81}; therefore, the high ejecta velocities in Type Ib/c supernovae lead us to expect little pollution of the companion.
Indeed, we find negligible pollution of the companion star with SN material.
However, the converse---pollution of the SNR with material from the companion---can be appreciable.
A few $10^{-2} \MSun$ of hydrogen-rich material may be stripped from the companion by the passing shell in our simulations, which may be detectable as an asymmetry in the metallicity of the SNR on the side of the companion.

\begin{figure*}
\includegraphics[width=1.95\columnwidth]{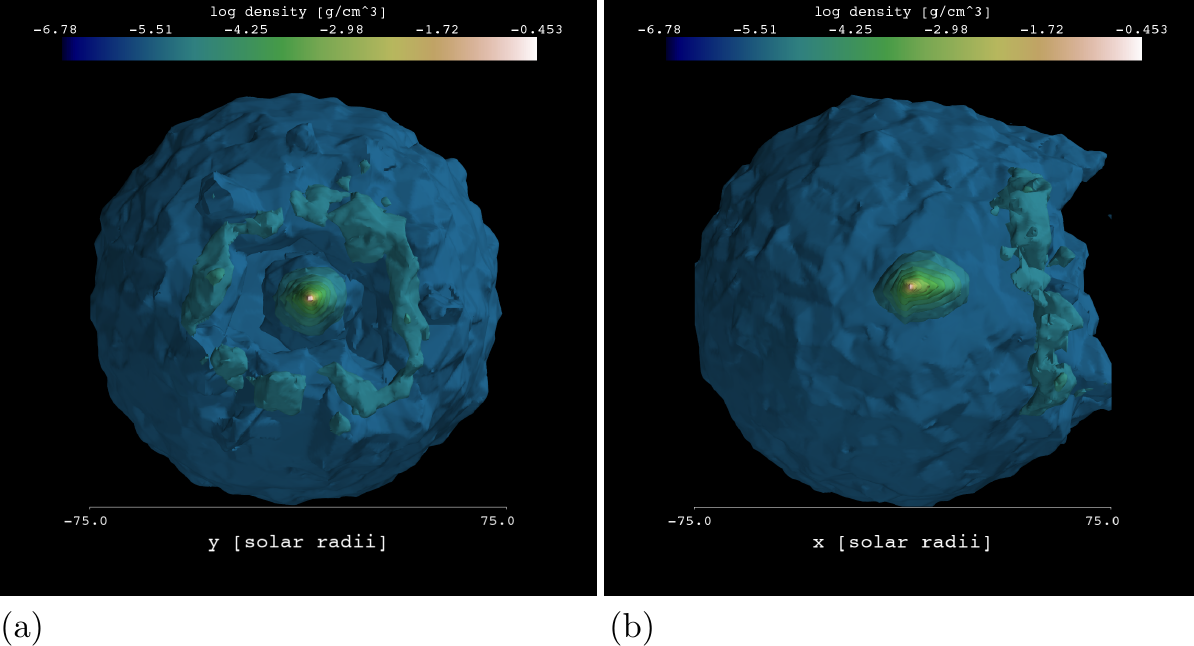}
\caption{A 3D rendering of the gas density in the system $10^3~\mathrm{s}$ after the SN, viewed down the $x$-axis (a; the original axis of the binary) and $y$-axis (b).
The companion is still distorted due to the impact, and has produced a hole in the expanding ejecta.
We used the software Mayavi2 \citep{MAYAVI} for the visualisation.
\label{fig:shell_3D}}
\end{figure*}

Figure \ref{fig:shell_3D} shows a 3D rendering of the SNR and companion at $10^{3}~\mathrm{s}$ after the moment of the SN.
At this point, a hole has been created in the passing shell due to the presence of the companion, which is seen to persist at later times.
The hole in the ejecta caused by the companion is approximately 30 degrees in size for the minimum orbital separations.

Even if an ejecta hole cannot be detected morphologically, the presence of a hole in SNR ejecta may allow the inference of a companion from a burst of UV radiation generated during the impact with the companion, which can escape through the less optically thick region of the companion shadow cone \citep{Kasen10}.
The hole may persist to late stages of the SNR despite some amount of refilling due to the subsequent rarefaction wave along with hydrodynamic instabilities \citep{Kasen10, Garcia-Senz12}.

Not only do we observe a hole in the SNR due to the companion star, but we also see an increase in the density in a ring surrounding the hole, as shown most clearly in Figure~\ref{fig:shell_3D}.
As shell material impacts the outer part of the companion star, where material is stripped and swept up with the ejecta, this ring of gas is also compressed in contrast with the freely expanding ejecta that do not interact with the companion. 
Aside from augmentation of the early SN light curve, our results also suggest that ring-like enhancements in density of the SNR could indicate the presence of a companion star.
Ring-like structures may easier to detect than a hole in the SNR as the enhancement in density may also be associated with an increase in radiative losses in the ring.

\begin{figure*}
\includegraphics[width=1.95\columnwidth]{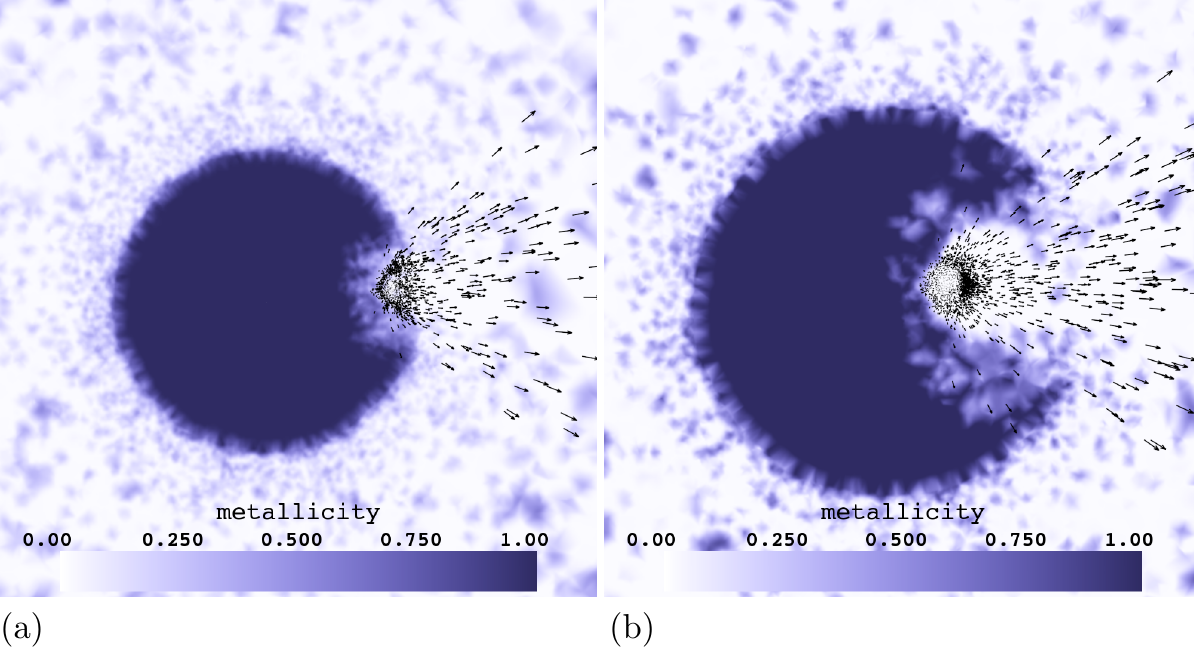}
\caption{A slice through the $x$-$y$ (orbital) plane showing the mean metallicity ($1 - X_\mathrm{H} - X_\mathrm{He}$) within $1~\RSun$ of the plane at (a) $700~\mathrm{s}$ and (b) $2000~\mathrm{s}$ after the SN. Black vectors show samples of the momentum of the material unbound from the companion star projected onto the orbital plane, as in Figure~\ref{fig:companion_snapshots}.
\label{fig:metallicity}}
\end{figure*}

Figure \ref{fig:metallicity} illustrates that the metallicity of the SNR is highest in the innermost regions, where the ejecta represents material nearest the core of the supernova progenitor.
A hole develops in this high-metallicity ejecta, at first primarily due to the shadow of the companion star (panel a).
At later times (panel b), the ablation of companion material further reduces the metallicity of a large fraction of the inner part of SNR in the direction of the companion.
The orbital motion of the companion star within the inner SNR during this longer period of ablation can also enlarge the region over which the gas is enriched with hydrogen.

\subsection{Post-impact evolution of the companion} \label{sec:results_evolution}
Following the stripping and ablation of mass from the outer layers of the companion star, we use \AMUSE\ to model the stellar evolution of the companion and compare with an unperturbed stellar model evolving from the main-sequence.
As we associate composition with each SPH particle from the original stellar model, we can convert the final SPH state of the companion back to a 1D structure model by an inversion of the method to construct the SPH model outlined in Section~\ref{sec:method_hydro_setup}.
After the 1D model is loaded back into \MESA, we continue the stellar evolution and compare the final age and HR diagram tracks to an undisturbed companion model.

A $1~\MSun$ star with metallicity $Z = 0.02$ evolves to the through to a carbon-oxygen WD at $12.1~\mathrm{Gyr}$ in \MESA.
On the other hand, the $1~\MSun$ model which has lost $0.04~\MSun$ of material from the SN impact reaches this stage at a later age of $14.0~\mathrm{Gyr}$.
Both the primary and companion stellar models were evolved for the same length of time (that is, the time it took for the primary star to reach its helium star stage).

\begin{figure}
\includegraphics[width=0.95\columnwidth]{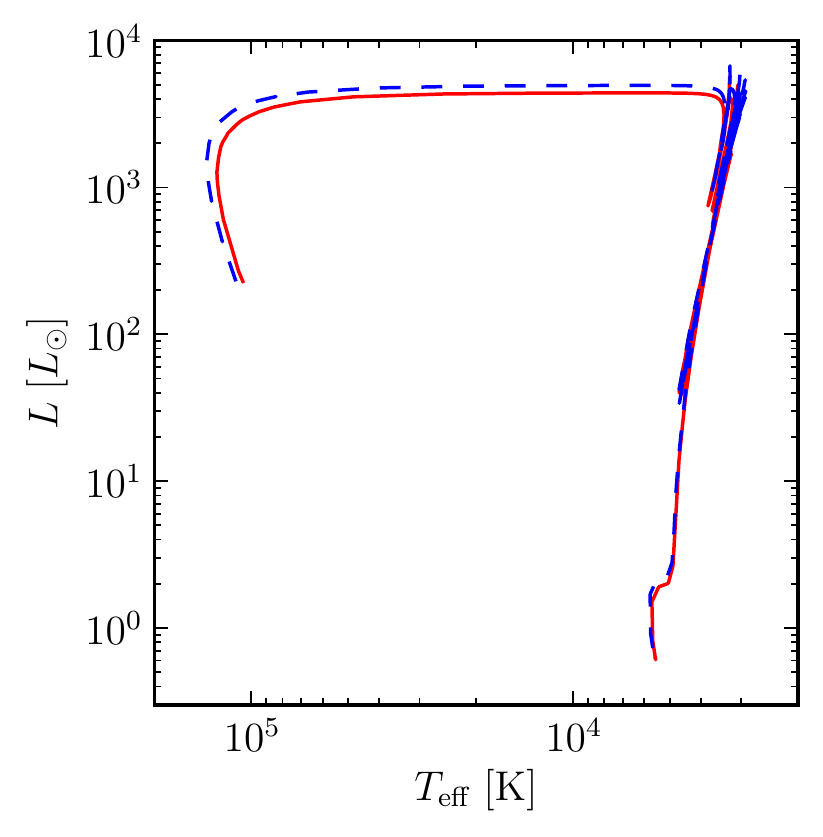}
\caption{Hertzsprung--Russell diagram of the evolution of the companion star with (red solid line) and without (blue dashed line) the impact of the supernova shell.
Tracks are plotted through to the oxygen/neon WD phase.
Beginning on the main sequence (from $t = 19.1~\mathrm{Myr}$), the model with the stripped envelope runs until reaching the carbon-oxygen WD phase at $14.0~\mathrm{Gyr}$, while the reference model (unstripped envelope) runs until the carbon-oxygen WD phase at $12.1~\mathrm{Gyr}$.
\label{fig:HR}}
\end{figure}

Although the final age of the stars is noticeably different, there is not a large evolutionary difference as represented on the HR diagram of Figure \ref{fig:HR}.
The luminosity of the stripped star is somewhat lower during its evolution, but in general this discrepancy is only around $20\%$ of the value seen in the undisturbed companion model.
It may therefore be difficult to distinguish a companion that has lost part of its envelope due to a supernova from $T_\mathrm{eff}$ and $L$ alone.
However, the stripping and contamination in the outer layers of the star might produce differences in chemical abundances that are spectrally distinguishable from the coeval stellar population \citep[see also, for example,][]{Pan12}.

\section{Discussion and conclusions} \label{sec:conclusions}
For supernovae in close binaries, the impact of the ejecta shell can have non-negligible effects on the mass and velocity of the companion star.
This is therefore an important phenomenon in considerations of the runaway velocities from supernova-disrupted binaries and, for example, their potential to contaminate searches for hypervelocity stars from other origins, such as the Hills mechanism with the supermassive black hole in the Galactic Centre \citep{Hills88, Yu03}.

We have performed SPH simulations of SNe in close binaries to study the consequences of the shell impact on the companion.
The overall hydrodynamic phenomena and trends we observe during these simulations are broadly consistent with previous studies of Type Ia \citep{Marietta00, Pakmor08, Pan12, Liu12}, Type Ib/c \citep{Liu15} and Type II \citep{Hirai14} SNe.
In addition we find that the gradient in the impact velocity predicted by \cite{Wheeler75} matches our results well, with some modification of the $\eta$ parameter representing the total momentum received by the companion.

While this work was in preparation, \cite{Liu15} presented work on the effect of a Type Ib/c supernova shell impacting onto a companion star, in order to derive the linear momentum transfer between the shell and star.
The velocity induced onto the companion due to the shell impact in their work is a factor of $1.5 \sim 2$ lower than our results.
Although it is not straightforward to separate the causes of these discrepancies, there are a number of differences between our calculations.
Most notably, in our case, the shell is naturally formed from the supernova explosion mechanism, that originated in the exploding star, as opposed to the introduction of the supernova shell by an analytic description of the material that impacts the companion.

\begin{figure}
\includegraphics[width=0.95\columnwidth]{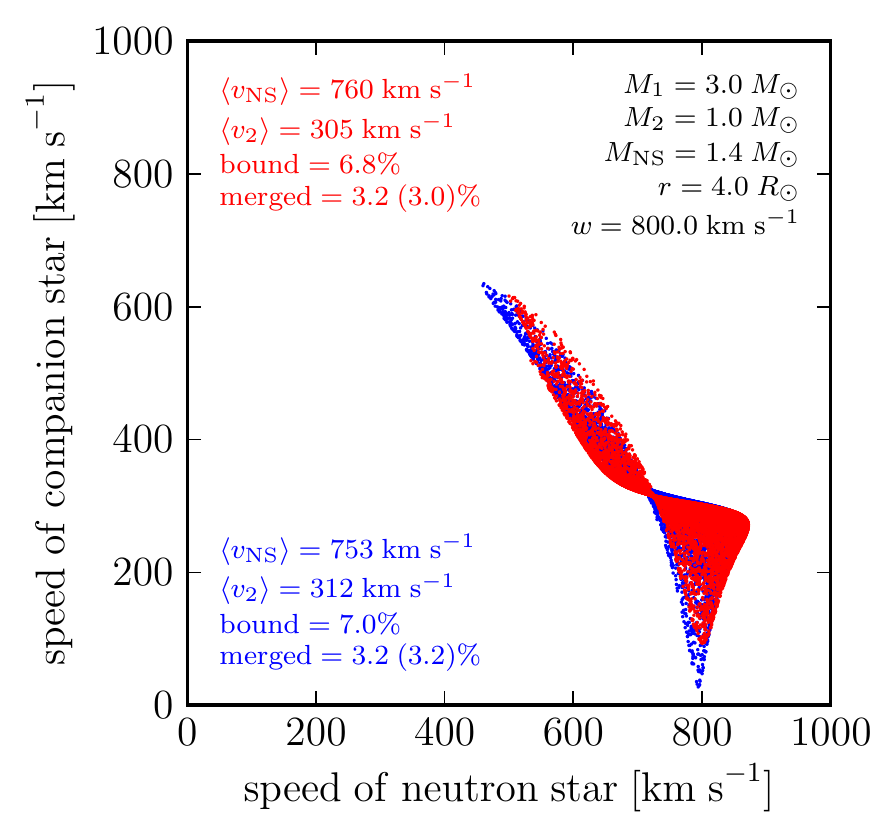}
\caption{Comparison of Monte Carlo sampling over NS kick orientation for the case with no impact effects on the companion (blue) and with impact effects as calibrated from our simulations (red).
Both cases show $10^{4}$ samples of NS kick orientation with a uniform on distribution over the unit sphere.
The magnitude of the NS kick velocity, $w$, is fixed at $800~\mathrm{km\,s}^{-1}$ throughout.
Mean values of the runaway velocities, as well as percentages of cases where the binary components remain bound and merged (and, in parentheses, merged cases that were calculated as bound).
Neither bound nor merged cases appear in these distributions.
\label{fig:MC}}
\end{figure}

Using the calibration from our simulations, we return to the question of runaway velocities of the components of supernova-disrupted binaries considered in TT98 and \cite{Tauris15}.
We have created a \texttt{python} code that calculates the final speeds derived by TT98 in order to investigate the analytic predictions with our simulation results.
In this Monte Carlo code, an impulsive increase in velocity, $\vect{w}$, is imposed to the neutron star, randomly oriented from an isotropic distribution over a sphere.
This is achieved by mapping from a uniform random distribution over $t \in (0, 1]$ to $2 \pi t$ for the angle $\phi$, and from a uniform random distribution over $u \in [0, 1]$ to $\cos^{-1}(2u - 1)$ for the angle $\theta$.
Figure \ref{fig:MC} shows a comparison of the distributions of speeds with (red) and without (blue) the effect of applying $\vect{v_\mathrm{im}}$ and mass loss in the companion star.

\begin{figure*}
\includegraphics[width=1.95\columnwidth]{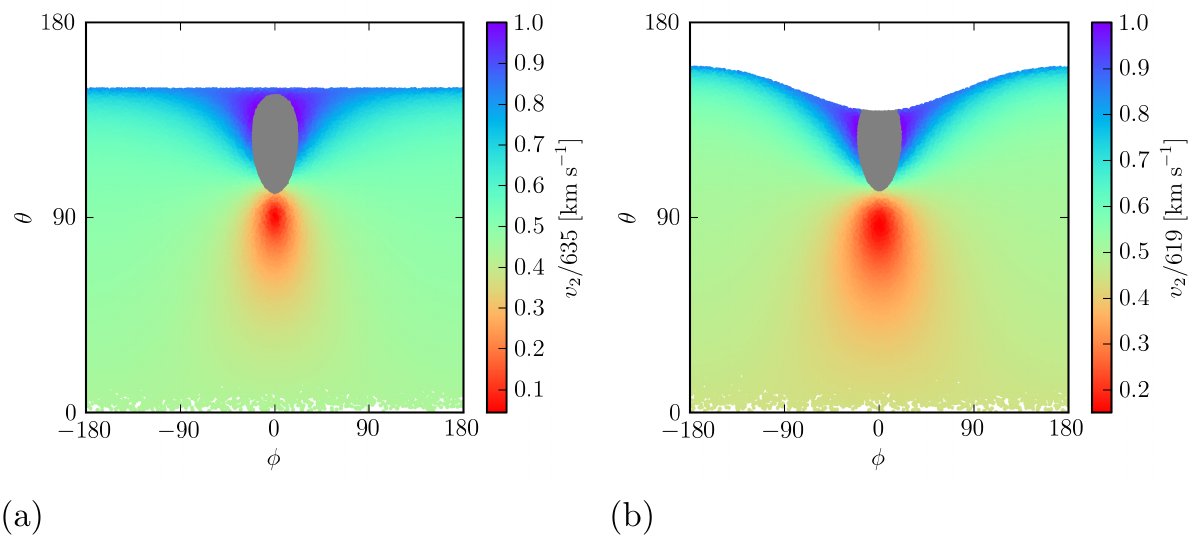}
\caption{Distribution of NS kick angles from the $x$-axis (aligned with the pre-kick NS orbital vector) for the same parameters as the Monte Carlo runs shown in Figure~\ref{fig:MC} (but with $2 \times 10^{5}$ samples to better fill the space in angles).
The span of $\theta$ over $\phi = 0$ defines the orbital plane, where $\theta = 0$ is for a kick aligned with the NS orbital velocity vector.
Panel (a) shows the case of no impact effects on the companion (blue distribution in Figure~\ref{fig:MC}) and panel (b) shows the case where impact effects calibrated from our simulations are included (red distribution in Figure~\ref{fig:MC}).
Colours represent the magnitude of the companion star from disrupted binaries (as a fraction of the maximum runaway velocity).
Grey shows cases where the NS and companion star merge.
The white regions for large $\theta$ are cases where the binary remains bound and so there is no runaway companion.
\label{fig:angles}}
\end{figure*}

From Figure~\ref{fig:MC}, it is evident that, although adding an impact velocity to the companion (perpendicular to its orbital velocity) increases the minimum companion speed, it also in fact reduces the maximum companion speed.
To clarify the discrepancies in the distributions that occur when adding $\vect{v_\mathrm{im}}$, we consider the effect of NS kick angles on the final velocity of the companion star in disrupted binaries in Figure~\ref{fig:angles}, analogous to Figure~4 in \cite{Tauris15}.
The white regions for high $\theta$ in each panel represent binaries that remain bound after the NS kick (and thus the runaway velocity is undefined).
The grey regions represent NS kick angles for which the NS and companion star merge after the SN.
It can be seen from the lower panel that the effect of applying an impact velocity to the companion star can stabilise the systems where the NS kick is counter-aligned with the NS orbital velocity.
In fact, the small region of parameter space giving large values of $v_2$ at $\phi = 0$ and high $\theta$ is removed after adding $\vect{v_\mathrm{im}}$ (due to these systems now remaining bound).
This explains the potentially counter-intuitive result that by adding an additional velocity to the companion star in fact reduces the maximum possible velocity of runaway stars.

The main results of our work are as follows:
\begin{enumerate}
\item  We follow the supernova self-consistently from just after the core bounce in a helium star generated from a stellar evolution model. Exploding a SN in such a model produces an ejecta profile that is different from that used in previous work, which has employed an analytic function for the ejecta distribution derived from the theory of shocks travelling through a one-dimensional atmosphere.
The progenitor model used here is still somewhat artificial in construction, with a constant mass loss parameter late in its evolution.
As understanding of Type Ib/c progenitors improves \citep[for some recent investigations, see][]{Kim15}, future work would benefit from a more realistic progenitor model by modelling the mass loss processes in detail.
\item We have investigated the mass removed from the companion in the very close binary separations seen in Type Ib/c SNe, as well as the net change in momentum of the companion star due to the shell impact and later ablation of material.
The change in momentum of the companion is used to calibrate theoretical predictions of the final velocity components in dissociated binaries, which is useful for studies of runaway stars from SNe, as seen in the recent work of \cite{Tauris15}.
\item We investigated the morphology of the SNR shortly after the shell has passed the companion, as well as the pollution of the SNR with material stripped from the companion, which, for the case of Type Ib/c SNe, may be a detectable fraction of the total mass in the ejecta (several $10^{-2}~\MSun$ out of $\sim 2 \MSun$).
The metallicity of the SNR is found to be highest in the inner regions of the SNR, behind the expanding shell at the shock front, and in this region the ablation of hydrogen from the outer layers of the companion star can dilute the metallicity on the side of the SNR facing the companion, resulting in a strong asymmetry in metallicity in the orbital plane.
\item The companion star is additionally found to modify the morphology of the SNR in two distinct ways: as anticipated, a hole forms in the SNR on the side of the companion; also, an increase in the SNR density is seen in a ring around the hole, which may enhance the luminosity in SNR observations.
\item We have also followed the stellar evolution of the companion star after the removal of mass during the shell impact, and have shown that the (main-sequence) lifetime of a star that has suffered a supernova impact can be substantially lengthened while traversing a slightly different track on the HR diagram.
\end{enumerate}

\begin{backmatter}

\section*{Competing interests}
The authors declare that they have no competing interests.

\section*{Author's contributions}
AR ran the simulations, analysed the data and wrote the draft of this paper.
SPZ had the idea that initiated this work and assisted with the interpretation of the results.
EMR assisted with the interpretation of the impact momenta and velocity results and the runaway star Monte Carlo results.
All of the authors took part in discussions concerning the results and contributed corrections and improvements on the draft of the manuscript.
All authors read and approved the final manuscript.

\section*{Acknowledgements}
We are grateful to Nathan de Vries for his assistance with an early version of the supernova explosion code in \AMUSE\, and to Arjen van Elteren and Inti Pelupessy for \AMUSE\ code development.
This work was supported by the Netherlands Research Council (NWO grant numbers 612.071.305 [LGM] and 639.073.803 [VICI]) and by the Netherlands Research School for Astronomy (NOVA).


\bibliographystyle{bmc-mathphys} 
\bibliography{binary_sn_simulations} 


\end{backmatter}
\end{document}